\documentclass[floatfix]{revtex4}
\usepackage{amsmath}
\usepackage{graphicx}

\begin{document}

\title{A new effective-field technique for the ferromagnetic spin-1
Blume-Capel model in a transverse crystal field}
\author{J. Roberto Viana$^{a}$}
\author{Octavio D. Rodriguez Salmon$^{a}$}
\author{Minos A. Neto$^{a}$}
\author{Diego C. Carvalho$^{b}$}
\affiliation{$^{a}$ Departamento de F\'{\i}sica, Universidade Federal do Amazonas, 3000,
Japiim, 69077-000, Manaus-AM, Brazil\\
$^{b}$Instituto Federal Norte de Minas Gerais - Campus Salinas,
MG - CEP:39560-000 , Brazil}
\date{\today}

\begin{abstract}
A new approximating technique is developed so as to study the quantum
ferromagnetic spin-1 Blume-Capel model in the presence of a transverse
crystal field in the square lattice. Our proposal consists of approaching
the spin system by considering islands of finite clusters whose frontiers
are surrounded by non-interacting spins that are treated by the
effective-field theory. The resulting phase diagram is qualitatively
correct, in contrast to most effective-field treatments, in which the
first-order line exhibits spurious behavior by not being perpendicular to
the anisotropy axis at low temperatures. The effect of the transverse
anisotropy is also verified by the presence of quantum phase transitions.
The possibility of using larger sizes constitutes an advantage to other
approaches where the implementation of larger sizes is costly computationally.

\end{abstract}

\maketitle

\section{Introduction}

In general, many-body systems with interactions are very difficult to solve
exactly. A way to overcome this difficulty is by approaching the many-body
problem by a one-body problem, in which a mean-field replaces the
interactions affecting the body. This idea is applied to the ferromagnetic
Ising Model (see reference \cite{kadanoff1}). In the most simple mean-field
approach, the nearest-neighbor interactions affecting each spin $S_{i}$ are
replaced in such a way that $S_{i}$ now interacts with an effective field
given by $zJ \langle S_{i} \rangle$, where $z$ is the coordination number, $%
J $ the exchange constant, and $\langle S_{i} \rangle$ is the thermal
average of the spin $i$. This is the so called "Weiss mean-field approach" %
\cite{weiss}. Nevertheless, it neglects the spin correlations, and it leads
the transition temperature $T_{c}$ as well as the values of the critical
exponents away from the exact values ($T_{c} = zJ/k_{B}$, for all
dimensions). However, for the one-dimensional case, the Ising model lacks of
a phase transition at finite temperature, but Weiss' approach wrongly
predicts that $T_{c} = 2J/k_{B}$. A further step for improving the solution
of this problem, is to use the proposal of Hans Bethe, which consists in
considering that a central spin should interact with all its
nearest-neighbour spins forming a cluster \cite{bethe}. Then, that cluster
would interact to an effective field that approaches the
next-nearest-neighbor spins surrounding the cluster. Thus, this improvement
gives $T_{c}=2J/k_{B}\ln(z/(z-2))$, which not only betters the approximation
of the critical temperatures, but leads correctly to $T_{c}=0$, for the
one-dimensional case. In this way, the correlations between the spins has
been included to some degree by considering a cluster of spins interacting
with its nearest-neighbors.\newline

A further step in approaching the many-body problem in spin systems is the effective-field approach. It is used in spin models with finite-size clusters  based on the following Hamiltonian splitting:

\begin{equation}
\mathcal{H}=\mathcal{H}_{c}+\mathcal{H}_{v},
\end{equation}%
\noindent where $\mathcal{H}_{c}$ corresponds to the energy that is composed
of spin variables of the finite cluster, whereas $\mathcal{H}_{v}$,
corresponds to the energy of the neighborhood, whose spins do not belong to
the central sites of the finite cluster. In the canonical ensemble, the
calculation of mean values of the spin variables $G_{c}$ belongs to the
subspace $n_{c}$ of the finite cluster, and it is computed by the following
procedure:

\begin{align}
\left\langle G_{c}\right\rangle & =\frac{TrG_{c}\exp (-\beta \mathcal{H})}{%
Tr\exp (-\beta \mathcal{H})}  \notag \\
& =\left\langle \frac{Tr_{n_{c}}G_{c}\exp (-\beta \mathcal{H}_{c})}{%
Tr_{n_{c}}\exp (-\beta \mathcal{H}_{c})}\right\rangle ,  \label{MO}
\end{align}%
\vskip \baselineskip This equation is exact if $\left[ \mathcal{H}_{c},%
\mathcal{H}_{v}\right] =0$. The great merit of Eq.(\ref{MO}), is that we can
solve the model of the infinite system by using a finite system in the
subspace $n_{c}$. Various approximation methods use Eq.(\ref{MO}) as a
starting point. One of them is the effective-field theory (EFT) proposed by
Honmura and Kaneyoshi \cite{Kaneyoshi} for solving the spin-1/2
ferromagnetic system. Sousa \textit{et al.} \cite%
{Jsousa1,Jsousa2,Jsousa3,Jsousa4} applied EFT so as to treat different
magnetic models with competing interactions. Recently, Viana \textit{et al.} %
\cite{Viana} developed a mean-field proposal for spin models, denominated
effective correlated mean-field (ECMF), based on the following ansatz:

\begin{equation}
\sigma _{j}=\lambda \left\langle S_{c}\right\rangle ,  \label{REL3}
\end{equation}%
\vskip\baselineskip \noindent where $\sigma _{j}$ are the neighbors of the
central spins of the finite cluster, $\left\langle S_{c}\right\rangle $ is
the mean of the spin variable of the cluster, and $\lambda $ is a term
exhibiting the behavior of a molecular parameter.

The aim of this proposal is the improvement of the results obtained by other
effective-field techniques, but we believe that the main advantage of our
proposal is the simplicity in treating first-order phase transitions.
Accordingly, in this work, we test our new technique in a quantum version of
the spin-1 Blume-Capel model in the presence of a transverse crystal field
in the square lattice. \newline

We remark that the classical Blume-Capel model ($\mathbf{BC}$) \cite%
{Blume,Capel} is one of the most suitable models for studying magnetic
systems from the point of view of the Statistical Mechanics. This model and
its generalization, the Blume-Emery-Griffiths model, ($\mathbf{BEG}$) was
proposed to describe the $\lambda $ transition in $\mathbf{^{4}He-^{3}He}$
mixtures \cite{beg} as well as ordering in a binary alloy \cite{rys,hint}.
Furthermore, its applications also include the description of ternary fluids %
\cite{mukamel1974,furman1977}, solide-liquid-gas mixtures and binary fluids %
\cite{laj1975,siv1975}, microemulsions \cite{schick1986,gompper1990},
ordering in semiconducting alloys \cite{newman1983,newman1991} and electron
conduction models \cite{kivelson1990}. Indeed, the BC model is found in many
works using different lattices, spin degrees, including disorder and
different Statistical Mechanic techniques \cite%
{Grollau,Kimel,Xavier,Plascak,zaim2008,Polat1,Polat2,Zhang,Kopec}. \newline

The Hamiltonian of the original BC model is as follows:

\begin{equation}
\mathcal{H}_{N}=-J\overset{N}{\underset{i\neq j}{\sum }}%
S_{i}^{z}S_{j}^{z}+D_{z}\overset{N}{\underset{i=1}{\sum }}\left(
S_{i}^{z}\right) ^{2},
\end{equation}%
\vskip\baselineskip \noindent where $J$ is the ferromagnetic coupling
between the spins $S_{i}^{z}= 0, \pm 1$ of the lattice and $D_{z}$ is the
anisotropy parameter. At zero temperature, for $0 < D_{z} < D_{c}$, the
energy of the Hamiltonian $\mathcal{H}_{N}$ is minimized when all spins are $%
S_{i}^{z}=\pm 1$, but for $D_{z}> D_{c}$, all the spins take the value $%
S_{i}^{z}=0$, so the system suffers a first-order phase transition at $%
D=D_{c}$. The critical value $D_{c}$ can be determined by equating the
energy of the order and disordered states, i.e.,

\begin{equation}
\mathcal{H}_{N}(S_{j}^{z}=\pm 1)=\mathcal{H}_{N}(S_{j}^{z}=0).
\end{equation}
\vskip \baselineskip

When the temperature is taken into account, the BC model provides us a phase
diagram with a a tricritical point separating a second and a first-order
frontier that divides the ferromagnetic order (F) and the paramagnetic
region (PM). This rich critical behavior qualifies the BC model for
representing different phase transitions. The phase diagram of the BC model
with equivalent-neighbor interactions can be seen in Fig.2 of reference \cite%
{octavio}, which is qualitatively the same for dimensions greater than one.

A variant of this model is the Biaxial Blume-Capel model that considers a
transverse crystal or anisotropy field to the easy axis of magnetization:

\begin{equation}
\mathcal{H}_{N}=-J\underset{i\neq j}{\sum }S_{i}^{z}S_{j}^{z}+\overset{N}{%
\underset{j=1}{\sum }}\left[ -D_{x}\left( S_{j}^{x}\right) ^{2}+D_{z}\left(
S_{j}^{z}\right) ^{2}\right] .  \label{BCT1}
\end{equation}%
Thus, the anisotropies $D_{x}$ e $D_{z}$ are now relevant physical
parameters that play an important role in the quantum phase transitions that
emerge. So, the term containing $D_{x}$ enriches the phase diagram of the BC
model. \newline
In classical phase transitions the spins are oriented according to
temperature fluctuations. On the other hand, the quantum phase transitions
occur at very low temperatures, so the spins are oriented by quantum
fluctuations associated to states of energy minima. Particularly, in this
model there are critical values of $D_{x}$ e $D_{z}$ for which
ferromagnetic-paramagnetic phase transitions exist for $T\simeq 0$.

When $D_{x}=0$ and $D_{z}\neq 0$ a quantum phase transition (at $T=0$)
exists for the critical value of $D_{z}$ corresponding to an energy minimum
of zero value related to the state $S_{j}^{z}=0$. However, for $D_{z}=0$ and 
$D_{x}\neq 0$, the quantum phase transition occur for a critical value of $%
D_{x}$ related to a energy minimum $-D_{x}$ related to the state where $%
S_{j}^{z}=0$.

\bigskip

\section{Implementation of the technique}

In the present proposal the system consists of finite-size clusters of
interacting sites $\mathbf{S}_{j}=\mathbf{S}_{j}\left(
S_{j}^{x},S_{j}^{y},S_{j}^{z}\right)$ and non-interacting sites $\sigma _{j}$%
, which belong to a set of inifine N particles. In this work we use the
cluster scheme shown in Fig.1, where we may observe islands of finite size $%
\mathbf{C}_{w}$ that are composed by $N_{c}$ interacting sites $\mathbf{S}%
_{w,j}$, in which different $\mathbf{C}_{w}$ clusters do not interact, and
spins $\sigma _{w}$ have only $z$-component. \newline
The Hamiltonian of the model given in Eq.(\ref{BCT1}) is, regarding the
scheme of Fig.1, as follows:

\begin{equation}
\mathcal{H}_{N}=\mathcal{H}_{\sigma }+\mathcal{H}_{w},  \label{BCT2}
\end{equation}%
where 
\begin{equation}
\mathcal{H}_{\sigma }=-J\overset{n}{\underset{j=1}{\sum }}\overset{4}{%
\underset{r=1}{\sum }}\sigma _{j}S_{j,r}^{z}+D_{z}\overset{n}{\underset{j=1}{%
\sum }}\sigma _{j}^{2},  \label{Hs}
\end{equation}%
that corresponds to the portion of the energy of spins $\sigma _{j} $.
Furthermore, we have

\begin{equation}
\mathcal{H}_{w}=\underset{w}{\sum }\left\{ -J\overset{N_{w}}{\underset{i\neq
j}{\sum }}S_{w,i}^{z}S_{w,j}^{z}+\overset{N_{w}}{\underset{j=1}{\sum }}\left[
-D_{x}\left( S_{w,j}^{x}\right) ^{2}+D_{z}\left( S_{w,j}^{z}\right) ^{2}%
\right] \right\} ,  \label{Hw}
\end{equation}%
which is the portion of the energy that excludes spins $\sigma _{j}$. Thus,
we have the following equations associated to the term $-\beta \mathcal{H}%
_{N}$ :

\begin{equation}
-\beta \mathcal{H}_{\sigma }=K\overset{n}{\underset{j=1}{\sum }}\overset{4}{%
\underset{r=1}{\sum }}\sigma _{j}S_{j,r}^{z}-Kd_{z}\overset{n}{\underset{j=1}%
{\sum }}\sigma _{j}^{2}
\end{equation}%
and also

\begin{equation}
-\beta \mathcal{H}_{w}=\underset{w}{\sum }\left\{ K\overset{N_{w}}{\underset{%
i\neq j}{\sum }}S_{w,i}^{z}S_{w,j}^{z}+\overset{N_{w}}{\underset{j=1}{\sum }}%
\left[ Kd_{x}\left( S_{w,j}^{x}\right) ^{2}-Kd_{z}\left( S_{w,j}^{z}\right)
^{2}\right] \right\} ,
\end{equation}%
where $\beta =1/k_{B}T,$ $d_{x}=D_{x}/J$ e $K=\beta J$. \ \ 

Now we have to perform the mean in the $v$ space of the spins $\sigma _{j}$,
given by : 
\begin{eqnarray*}
\left\langle \left\langle \sigma _{j}\right\rangle _{v}\right\rangle
&=&\left\langle \frac{Tr_{v}\sigma _{j}\exp \left( -\beta \mathcal{H}%
_{N}\right) }{Tr_{v}\exp \left( -\beta \mathcal{H}_{N}\right) }\right\rangle
\\
&=&\left\langle \frac{Tr_{v}\sigma _{j}\exp \left( -\beta \mathcal{H}%
_{\sigma }\right) }{Tr_{v}\exp \left( -\beta \mathcal{H}_{\sigma }\right) }%
\right\rangle \\
&=&\left\langle \frac{\partial }{\partial \phi _{j}}\ln \left( Z_{j}\right)
\right\rangle
\end{eqnarray*}%
where 
\begin{equation*}
Z_{j}=Tr_{v}\exp \left( -\beta \mathcal{H}_{\sigma }\right) =2\exp \left(
-Kd_{z}\right) \cosh \left( \phi _{j}\right) +1, 
\end{equation*}%
and also%
\begin{equation*}
\phi _{j}=K\overset{4}{\underset{r=1}{\sum }}S_{j,r}^{z}. 
\end{equation*}%
In this way we have 
\begin{equation*}
\left\langle \left\langle \sigma _{j}\right\rangle _{v}\right\rangle =\frac{%
2\exp \left( -Kd_{z}\right) \sinh \left( \phi _{j}\right) }{2\exp \left(
-Kd_{z}\right) \cosh \left( \phi _{j}\right) +1}=\left\langle G\left( \phi
_{j}\right) \right\rangle 
\end{equation*}%
Now, we apply the differential operator technique by regarding $%
S_{j,r}^{z}=0,\pm 1$ \'{e}, obtaining the following result: 
\begin{eqnarray}
\left\langle \left\langle \sigma _{j}\right\rangle _{v}\right\rangle
&=&\left\langle \exp \left( K\overset{4}{\underset{r=1}{\sum }}S_{j,r}^{z}%
\frac{\partial }{\partial x}\right) G_{j}\left( x\right) \left| 
\begin{array}{c}
\\ 
x=0%
\end{array}%
\right. \right\rangle  \notag \\
&=&\left\langle \overset{4}{\underset{r=1}{\prod }}\exp \left( KS_{j,r}^{z}%
\frac{\partial }{\partial x}\right) G_{j}\left( x\right) \left| 
\begin{array}{c}
\\ 
x=0%
\end{array}%
\right. \right\rangle  \notag \\
&=&\left\langle \overset{4}{\underset{r=1}{\prod }}\left[ \overset{2}{%
\underset{p=0}{\sum }}b_{p}\left( S_{j,r}^{z}\right) ^{p}\right] G_{j}\left(
x\right) \left| 
\begin{array}{c}
\\ 
x=0%
\end{array}%
\right. \right\rangle  \label{media1}
\end{eqnarray}%
where $\frac{\partial }{\partial x}$ is the differential operator, and also%
\begin{equation}
b_{0}=1,\text{ }b_{\text{1}}=\sinh \left( K\frac{\partial }{\partial x}%
\right) \text{ e }b_{\text{2}}=\cosh \left( K\frac{\partial }{\partial x}%
\right) -1.
\end{equation}

It is important to note that in this calculation process we have many
correlation means of sites $S_{j,r}^{z}$, however, sites $S_{j,r}^{z}$
belong to different clusters $C_{w}$, thus sites $S_{j,r}^{z}$ are
independent. Accordingly, the following first-order relations of approach
can be used by regarding the definition of the magnetization $m$:

\begin{eqnarray}
\left\langle S_{j,r}^{z}\right\rangle &=&m\text{ } \\
\left\langle S_{j,r_{1}}^{z}S_{j,r2}^{z}\right\rangle &=&\left\langle
S_{j,r_{1}}^{z}\right\rangle \left\langle S_{j,r2}^{z}\right\rangle =m^{2}%
\text{ } \\
\left\langle S_{j,r_{1}}^{z}S_{j,r2}^{z}S_{j,r3}^{z}\right\rangle
&=&\left\langle S_{j,r_{1}}^{z}\right\rangle \left\langle
S_{j,r2}^{z}\right\rangle \left\langle S_{j,r3}^{z}\right\rangle =m^{3} \\
\left\langle
S_{j,r_{1}}^{z}S_{j,r2}^{z}S_{j,r3}^{z}S_{j,r4}^{z}\right\rangle
&=&\left\langle S_{j,r_{1}}^{z}\right\rangle \left\langle
S_{j,r2}^{z}\right\rangle \left\langle S_{j,r3}^{z}\right\rangle
\left\langle S_{j,r4}^{z}\right\rangle =m^{4}.
\end{eqnarray}%
This treatment also applies for the terms e $\left\langle \left(
S_{j,r}^{z}\right) ^{2}\right\rangle =q$. So, Eq.(\ref{media1}) can be
rewritten in the following form: 
\begin{equation}
\left\langle \left\langle \sigma _{j}\right\rangle _{v}\right\rangle =\left[
b_{0}+b_{1}m+b_{2}q\right] ^{4}G_{j}\left( x\right) \left| 
\begin{array}{c}
\\ 
x=0%
\end{array}%
\right. .  \label{media2}
\end{equation}%
Now, by performing this trinomial operator we get the following result:%
\begin{equation}
\left\langle \left\langle \sigma _{j}\right\rangle _{v}\right\rangle =%
\overset{4}{\underset{p_{1}=0}{\sum }}\text{ }\overset{4-p_{1}}{\underset{%
p_{2}=0}{\sum }}a_{1}^{4-p_{1}-p_{2}}a_{2}^{p_{2}}a_{3}^{p_{1}}\exp \left[
\left( 4-p_{1}-2p_{2}\right) K\frac{\partial }{\partial x}\right]
G_{j}\left( x\right) \left| 
\begin{array}{c}
\\ 
x=0%
\end{array}%
\right. ,  \label{media3}
\end{equation}%
where 
\begin{equation}
a_{1}=\frac{1}{2}\left( q+m\right) \text{, }a_{2}=\frac{1}{2}\left(
q-m\right) \text{ e }a_{3}=1-q\text{.}
\end{equation}%
Now we apply the following relation 
\begin{equation}
\exp \left( \omega \frac{\partial }{\partial x}\right) G_{j}\left( x\right)
\left| 
\begin{array}{c}
\\ 
x=0%
\end{array}%
\right. =G_{j}\left( x=\omega \right) \text{,}
\end{equation}
so:%
\begin{equation}
\left\langle \left\langle \sigma _{j}\right\rangle _{v}\right\rangle =%
\overset{4}{\underset{p_{1}=0}{\sum }}\text{ }\overset{4-p_{1}}{\underset{%
p_{2}=0}{\sum }}a_{1}^{4-p_{1}-p_{2}}a_{2}^{p_{2}}a_{3}^{p_{1}}G_{j}\left(
x=4-p_{1}-2p_{2}\right) ,
\end{equation}%
and we have also that 
\begin{equation}
a_{1}^{e_{1}}a_{2}^{e_{2}}a_{3}^{e_{3}}=\left( \frac{1}{2}\right)
^{e_{1}+e_{2}}\overset{e_{1}}{\underset{t_{1}=0}{\sum }}\overset{e_{2}}{%
\underset{t_{2}=0}{\sum }}\overset{e_{3}}{\underset{t_{3}=0}{\sum }}%
(-1)^{t_{2}+t_{3}}q^{e_{1}+e_{2}-t_{1}-t_{2}+t_{3}}m^{t_{1}+t_{2}}.
\end{equation}%
In this way we have the following simplified expression: 
\begin{equation}
\left\langle \left\langle \sigma _{j}\right\rangle _{v}\right\rangle =%
\overset{4}{\underset{k=0}{\sum }}\text{ }A_{k}m^{k}.  \label{media4}
\end{equation}%
where $A_{k}=A_{k}\left( K,d_{z},q\right) $. We verified that $A_{k}$ is
zero for even values of $k$.\newline

There are many mean-field proposals that have been done so as to approach $%
\sigma _{j}$. In this paper we use the following relation: 
\begin{equation}
\sigma _{j}=\lambda m,  \label{sigma}
\end{equation}%
where $\lambda $ is a parameter to be determined. Then we apply Eq.(\ref%
{sigma}) in Eq.(\ref{media4}), which leads to the folliwing result: 
\begin{eqnarray}
\left\langle \left\langle \lambda m\right\rangle _{v}\right\rangle &=&%
\overset{4}{\underset{k=0}{\sum }}\text{ }A_{k}m^{k}  \notag \\
\lambda &=&\overset{4}{\underset{k=0}{\sum }}\text{ }A_{k}m^{k-1}.
\label{Lambda}
\end{eqnarray}%
Note that from this equation we have that $\lambda =\lambda \left(
K,d_{z},q,m\right) $.

\subsection{The Interacting Cluster}

In Fig.2 we may see that a $\mathbf{C}_{w}$ cluster contain spins $\mathbf{S}%
_{j}$, each of them interacting between next-nearest neighbors in the
respective finite-size square lattice. We can also observe that $\sigma _{k}$
represents the neighbors of the central sites $S_{j}^{z}$, which compose the
finite cluster of $N_{c}$ sites. Thus, the Hamiltonian is conveniently
written in the following form:

\begin{equation}
-\beta \mathcal{H}_{N_{c}}=K\underset{i\neq j}{\sum }S_{i}^{z}S_{j}^{z}+K%
\underset{j}{\sum }\left[ -d_{z}\left( S_{j}^{z}\right) ^{2}+d_{x}\left(
S_{j}^{x}\right) ^{2}\right] +\overset{N_{c}}{\underset{j=1}{\sum }}%
C_{j}S_{j}^{z}.  \label{HF}
\end{equation}%
\vskip\baselineskip\noindent where

\begin{equation}
C_{j}=K\overset{n_{j}}{\underset{k=1}{\sum }}\sigma _{k}.  \label{Cj}
\end{equation}%
\vskip \baselineskip In the present work the ansatz given in Eq. (\ref{REL3}%
) is our fundamental assumption, so

\begin{equation}
C_{j}=n_{j}\lambda K\left\langle S_{c}^{z}\right\rangle .  \label{Ch}
\end{equation}%
\vskip\baselineskip

In what follows  the size of the finite clusters is  considered  to be $N_{c}=1,$ $%
2, $ $4,$ $9,$ $25,36,$ $49\,\ $ with $64$ central sites. Thus, we have the
following relations:

\begin{align}
\text{ }N_{c}& =1\text{: }C_{j}=4\lambda K\left\langle
S_{c}^{z}\right\rangle .  \label{N1} \\
\text{ }N_{c}& =2\text{: }C_{j}=3\lambda K\left\langle
S_{c}^{z}\right\rangle .  \label{N} \\
\text{ }N_{c}& =4\text{: }C_{j}=2\lambda K\left\langle
S_{c}^{z}\right\rangle .  \label{N4} \\
\text{ }N_{c}& \geq 9\text{: }C_{j}=2\lambda K\left\langle
S_{c}^{z}\right\rangle \text{ or }C_{j}=\lambda K\left\langle
S_{c}^{z}\right\rangle \text{ or }C_{j}=0.\text{\ }  \label{N9}
\end{align}%
The magnetic properties per particle such as  $m=\left\langle S_{c}^{z}\right\rangle $
and $q=\left\langle \left( S_{c}^{z}\right) ^{2}\right\rangle $ are given by
the following statistical definitions:%
\begin{eqnarray}
\left\langle S_{c}^{z}\right\rangle &=&\frac{1}{N_{c}}\frac{Tr\left( \overset%
{N_{c}}{\underset{j=1}{\sum }}S_{j}^{z}\right) \exp (-\beta \mathcal{H}%
_{N_{c}})}{Z_{N_{c}}}  \label{MAG} \\
\left\langle \left( S_{c}^{z}\right) ^{2}\right\rangle &=&\frac{1}{N_{c}}%
\frac{Tr\left( \overset{N_{c}}{\underset{j=1}{\sum }}\left( S_{j}^{z}\right)
^{2}\right) \exp (-\beta \mathcal{H}_{N_{c}})}{Z_{N_{c}}},  \label{Q}
\end{eqnarray}%
where the partition function in the space of sites $S_{j}^{z}$ is given by

\begin{equation}
Z_{N_{c}}=Tr\exp (-\beta \mathcal{H}_{N_{c}})=\overset{3^{N_{c}}}{\underset{%
j=1}{\sum }}E_{j},
\end{equation}%
where $E_{j}$ are the eigenvalues of the Hamiltonian. Particularly, when $%
N_{c}=1$ the eigenstates $\bigskip $ $\left| s_{z}\right\rangle $ of the
orthogonal basis are given by :

\begin{equation}
\bigskip \left| 1\right\rangle =\left( 
\begin{array}{c}
1 \\ 
0 \\ 
0%
\end{array}%
\right) ,\text{ }\left| 0\right\rangle =\left( 
\begin{array}{c}
0 \\ 
1 \\ 
0%
\end{array}%
\right) \text{ e }\left| -1\right\rangle =\left( 
\begin{array}{c}
0 \\ 
0 \\ 
1%
\end{array}%
\right) .  \label{base}
\end{equation}%
The hamiltonian matrix $-\beta \mathcal{H}_{1}$ is given by 
\begin{equation}
-\beta \mathcal{H}_{1}=\left( 
\begin{array}{ccc}
C_{1}+Kd_{x}/2-Kd_{z} & 0 & Kd_{x}/2 \\ 
0 & Kd_{x} & 0 \\ 
Kd_{x}/2 & 0 & -C_{1}+Kd_{x}/2-Kd_{z}%
\end{array}%
\right) .  \label{MATRIX}
\end{equation}%
So eigenvalues obtained from the matrix $-\beta \mathcal{H}_{1}$ correspond
to the following expressions:%
\begin{eqnarray}
E_{1} &=&Kd_{x}  \label{E1} \\
E_{2,3} &=&Kd_{x}/2-Kd_{z}+\frac{1}{2}\sqrt{K^{2}d_{x}^{2}+4C_{1}^{2}.}
\label{E2}
\end{eqnarray}
while the eigenvectors are :

\begin{eqnarray}
\bigskip \left| E_{1}\right\rangle &=&\left( 
\begin{array}{c}
0 \\ 
1 \\ 
0%
\end{array}%
\right) =\text{ }\left| 0\right\rangle  \label{AE1} \\
\left| E_{2}\right\rangle &=&\left( 
\begin{array}{c}
1 \\ 
0 \\ 
R_{2}%
\end{array}%
\right) =\text{ }\left| 1\right\rangle +R_{2}\text{ }\left| -1\right\rangle
\label{AE2} \\
\left| E_{3}\right\rangle &=&\left( 
\begin{array}{c}
R_{3} \\ 
0 \\ 
1%
\end{array}%
\right) =\text{ }R_{3}\left| 1\right\rangle +\text{ }\left| -1\right\rangle
\label{AE3}
\end{eqnarray}%
where 
\begin{eqnarray}
R_{2} &=&\frac{Kd_{x}}{2C_{1}+\sqrt{K^{2}d_{x}^{2}+4C_{1}^{2}}} \\
R_{3} &=&-\frac{Kd_{x}}{2C_{1}+\sqrt{K^{2}d_{x}^{2}+4C_{1}^{2}}}
\end{eqnarray}%
\bigskip For $N_{c}>1$, eigenvectos and eigenvalues of $-\beta \mathcal{H}
_{N}$ \ are obtained by numerical methods.

\bigskip

An important issue is the thermodynamic treatment of the spin system,
accordingly, we use the free energy given by the following equation:

\begin{equation}
\phi =-\frac{1}{N_{c}}t\ln \left( Z_{N_{c}}\right) +\gamma m^{2},  \label{fi}
\end{equation}

\noindent where $t=k_{B}T/J$ is the reduced temperature and $\gamma $ is a
parameter to be determined. At the equilibrium, the free energy is
minimized, thus:

\begin{equation}
\frac{\partial \phi }{\partial m}=f_{m}\equiv 0\text{,}  \label{COND6}
\end{equation}%
where the function $f_{m}$ stands for the equation of state given by
\begin{equation}
f_{m}=m-\left\langle S_{c}^{z}\right\rangle ,  \label{FN3}
\end{equation}%
and we can determine the parameter $\gamma $ by using Eq. (\ref{COND6}).

\section{Results}

We firstly obtained the phase diagram of the BC model in the plane $t-d_{z}$
in a square lattice, based on the considerations of the previous section. In
Fig. 3 we show the phase diagram for clusters containing $N_{c}=1$ (in (a)), 
$N_{c}=4$ (in (b)) and $N_{c}=16$ (in (c)) central sites. We faced the
computational problem of solving the model for big clusters, inasmuch as the
number of accessible states corresponds to $3^{N_{c}}$ states. For instance,
for $N_{c}=16$ and $N_{c}=64$ sites, we have accessible states of order $10^{7}$ and $10^{30}$, respectively, which are huge numbers. Accordingly, for clusters of sizes $N_{c}>16$, we prefer only to calculate the critical
temperature $t_{c}$, for $d=0$, and the coordinates of the tricritical point 
$P(d_{t},t_{t})$. In this figure we may observe that the critical value of
the anisotropy corresponds to $d_{c}=2.0$, for $t\rightarrow 0$, which
agrees with exact results obtained when equating the energy of the ordered
state $(S_{j}^{z}=\pm 1)$, with the energy of the disordered one ($S_{j}^{z}=0$), for a finite system of $N$ sites, i.e.,

\begin{equation}
d_{c}=\frac{z}{2},
\end{equation}%
where $z$ is the coordination number of the lattice. The black circle
represents the tricritical point that separates the second-order and the
first-order frontier. We observe that the critical temperature $t_{c}$
decreases as the size of the cluster $N_{c}$ increases. The first-order
frontier correctly falls perperndicularly to the anisotropy axis, however,
in general, effective-field results do not reproduce this feature of the
first-order frontier (see the IEFT curve in Figure 5 of reference \cite%
{Polat}). \newline

\bigskip

In Table 1 we present the values of $t_{c}$, obtained for each cluster size $%
N_{c}$, where we compare the results of this work using ECMF with the
mean-field approximation that uses clusters (MFT), where $\lambda =1$, in
this case. We remark that Y\"{u}ksel \textit{et al.} \cite{Polat} obtained $%
t_{c}\simeq 1.690$, by using a Metropolis Monte Carlo simulation (SMC),
whereas Silva \textit{et al.} \cite{Plascak} obtained $t_{c}\simeq 1.714$
using Wang-Landau sampling. In references \cite{Beale} and \cite{Xavier} we
have $t_{c}\simeq 1.695$ and $t_{c}\simeq 1.681$, respectively. We may
observe that the results obtained by the ECMF approach are close to the SMC
values when the cluster size is increased. Nevertheless, if compared with
the MFT results, $t_{c}$ is better estimated by the ECMF method, regarding
the SMC results as a reference. \newline

The calculations of the tricritical points $P(d_{t},t_{t})$ through the ECMF
technique are shown in Table 2. We see that when the cluster size $N_{c}$ is
increased, the values of the critical anisotropy $d_{t}$ approach the values 
$d_{t}=1.966(2)$ and $d_{t}=1.974$, obtained by Monte Carlo simulations of
references \cite{Plascak} and \cite{Polat}, respectively. In what the value $%
t_{t}$ concerns, the convergence is closer to that of reference \cite{Polat}%
, which gives $t_{t}=0.56$, obtained by Monte Carlo simulations \cite{Polat}.

The coordinates of the tricritical point were determined through a Landau
expansion,

\begin{equation}
\phi (d,t)=\overset{\infty }{\underset{p=0}{\sum }}A_{p}(d,t)m^{p}.
\end{equation}%
From this equation, we are interested in solving the following system of
equations 
\begin{eqnarray}
A_{2}(d_{t},t_{t}) &=&0 \\
A_{4}(d_{t},t_{t}) &=&0,
\end{eqnarray}%
so as to obtain the tricritical point. Thus, we may note that 
\begin{equation}
A_{p}=-\frac{t}{N_{c}}\frac{1}{Z_{N_{c}}}\left( \frac{\partial ^{p}Z_{N_{c}}%
}{\partial m^{p}}\right) _{m=0}+\left( \frac{\partial ^{p}}{\partial m^{p}}%
\left( \gamma m^{2}\right) \right) _{m=0}
\end{equation}%
corresponds to the equation that determines the coefficients $A_{p}$. 
\newline

In Fig. 4 we show the phase diagram in the $t-d_{x}$ plane for $N_{c}=1$
(case (a)) and $N_{c}=2$ (case (b)). In Fig. 5 we show the detail of this
diagram for low temperature region. Both cases exhibit phase transitions of
first and second order, as well as the presence of two ticritical points,
respectively. For $N_{c}=1$ were obtained $P(d_{t}=11.93,t_{t}=2.10^{-4})$
and $P(d_{t}=6.67,t_{t}=0.99)$, whereas for $N_{c}=2$, we have $%
P(d_{t}=10.73,t_{t}=10^{-2})$ and $P(d_{t}=6.16,t_{t}=0.94)$. 

Another important aspect is shown in Fig. 6, for a cluster with $N_{c}=1$
within the ECMF approach and corresponds to Fig.5. There it is shown a
 free energy minimum at the transition point located at $P\left( t=0.001,d_{x}=11.842076\right) $. This is signaling a first-order phase transition point.  Three minima at the same level can be observed, one for $m=0$, and two symmetrical ones at $m=\pm m_{0}$, which
clearly identify a first-order phase transition due to coexisting phases. In this point the temperature is close to zero and the value of the free energy tends to $\phi=-E_{1}/K=-d_{x}$, which corresponds to the eigenvalue of the disordered state $\left| E_{1}\right\rangle $, where $m=0$. This is a quantum phase transition, without the influence of the temperature fluctuations. In Fig. 7 is shown the o behavior of the free energy minimum for the phase transition point $P\left( t=0.001,d_{x}=10.85502\right)$ using the ECMF approach with $N_{c}=2$ sites. This, of course, is a signal of  a second-order phase transition. \\\\

Recently, it has been studied the classical BC model using an effective-field technique (EFT) \cite{Emanuel} with $N_{c}=1$ site on the
square lattice. The same qualitative results can also be observed in reference \cite{Diego}. Similarly, in Fig. 8  we exhibit  the mean-field case of te present model $\left( \lambda =1\right) $, where we can see a tricritical point for $N_{c}=1$ (frontier line (a)), whereas, for $N_{c}=2$ (frontier line (b)), two tricritical points are present. Thus,  a  detailed criticality  of these frontiers are shown  in Fig.9 for the low temperature region, so as to observe that  the lower  tricritical point in line (b) is very close to the zero temperature. 

\section{Conclusions}

In this paper we study the ferromagnetic spin-1 Blume-Capel (BC) model with
nearest-neighbor interactions in the presence of a transverse crystal-field,
within a mean-field approach. We call this new approach as effective
correlated mean-field (ECMF). For the bidimensional case, we implemented the
model in the square lattice. The results show that when the size of the
cluster $N_{c}$ increases, the values of the critical temperature $t_{c}$
(for null anisotropy) tend to $1.690$, which is the Monte Carlo estimate of Y\"{u}ksel \textit{et al.} \cite{Polat} (see Table 1). Another important result is related to the estimate of the tricritical point $P(d_{t},t_{t})$ in comparison with the results of Silva \cite{Plascak} and Y\"{u}ksel\cite{Polat} (see Table 2). Our ECMF values of $d_{t}$ reasonably agree with the
results of the Monte Carlo simulations, as $N_{c}$ increases, whereas $t_{t}$ tends to the effective-field (EFT) result developed in reference \cite{Polat}. \newline

The results that consider the transverse crystal $D_{x}$ can be observed in
the phase diagram in the plane $t-d_{x}$. There we have the presence of
first- and second-order phase transitions for $D_{x}>0$, together with two
tricritical points, whereas for $D_{x}<0$, we only have a second-order
criticality. The evidence of a quantum phase transition is shown in Fig. 6,
where the energy minima is threefold degenerated with $\phi =-E_{1}/K$,
related to the eigenvalue of the disordered state $(m=0$), see Eq. (\ref{AE1}).\newline

The main merit of the ECMF approximation is the determination of the
molecular parameter $\lambda $, which is obtained from the effective-field
theory. Furthermore, the values of the critical temperature determined by
this approach (for a given value of $N_{c}$) converge faster in comparison
with other techniques like Monte Carlo \cite{Polat}, and when we compare them
with the results obtained by the usual mean field approach (see table 1).

\bigskip

On the other hand, the possibility of working with larger sizes constitutes
an advantage for analyzing finite-size effects. Finally, we hope that this
new  technique of approach can be applied satisfactorily in other models and
lattices.

textbf{ACKNOWLEDGEMENT}

This work was partially supported by CNPq and FAPEAM (Brazilian Research Agencies).

\vskip \baselineskip  
\begin{table}[tbp] \centering%
\caption{Critical temperatures obtained for various cluster sizes using the ECMF and MFT
techniques.\label{TABELA1}} 
\begin{tabular}{cccccccccc}
\hline\hline
Technique/$N_{c}$ & 1 & 2 & 4 & 9 & 16 & 25 & 36 & 49 & 64 \\ \hline\hline
\multicolumn{1}{l}{\ \ \ \ \ \textbf{ECMF}} & $2.468$ & $2.420$ & $2.342$ & $%
2.277$ & $2.241$ & $2.216$ & $2.108$ & $1.972$ & $1.914$ \\ 
\multicolumn{1}{l}{\ \ \ \ \ \textbf{MFT}} & $2.666$ & $2.552\ $ & $2.406$ & 
$2.309$ & $2.259$ & $2.236$ & $2.224$ & $2.145$ & $2.098$ \\ \hline\hline
\end{tabular}%
\end{table}%

\vskip \baselineskip

\begin{table}[tbp] \centering%
\caption{Values of tricritical points obtsined the MFT-MC approximation for several sizes the cluster..\label{TABELA2}} 
\begin{tabular}{cccccccccc}
\hline\hline
$N_{\mathbf{c}}$ & $1$ & $2$ & $4$ & $9$ & $16$ & $25$ & $36$ & $49$ & $64$
\\ \hline\hline
\multicolumn{1}{l}{\ \ \ \ $\ d_{t}$} & $1.848$ & $1.857$ & $1$.$872$ & $%
1.874$ & $1.876$ & $1.896$ & $1.916$ & $1.921$ & $1.935$ \\ 
\multicolumn{1}{l}{\ \ \ \ \ $t_{t}$} & $1.182$ & $1.164$ & $1.113$ & $1.095$
& $1.051$ & $1.002$ & $0.899$ & $0.835$ & $0.792$ \\ \hline\hline
\end{tabular}%
\end{table}%

\vskip\baselineskip

\bigskip 

\vskip \baselineskip

\begin{figure}[htp]
\begin{center}
\includegraphics[height=8.0 cm]{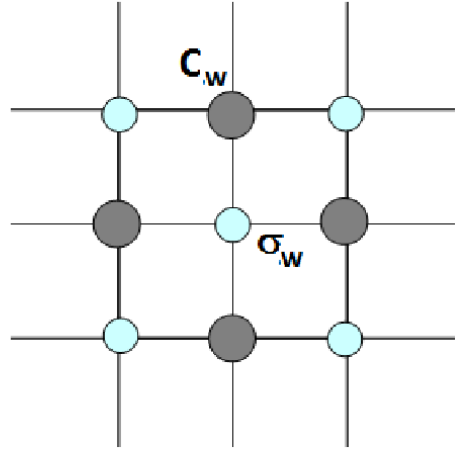}
\end{center}
\vspace{0.05cm}
\caption{ The scheme showing the cluster of spins to be used in the ECMF approach.}
\label{figura1} 
\end{figure}

\vskip \baselineskip
\begin{figure}[htp]
\begin{center}
\includegraphics[height=8.0 cm]{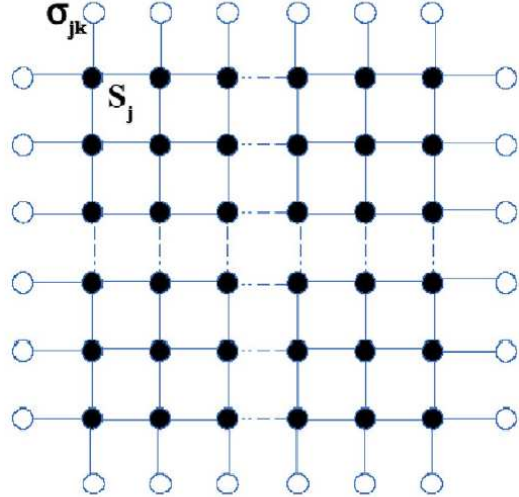}
\end{center}
\vspace{0.05cm}
\caption{ Scheme for sites located on a square lattice, where we have the
central sites (filled circles) and neighboring sites (open circles). }
\label{figura2} 
\end{figure}
\vskip \baselineskip
\vspace{3.0 cm}
\begin{figure}[htbp]
\centering
\includegraphics[height=9.0cm]{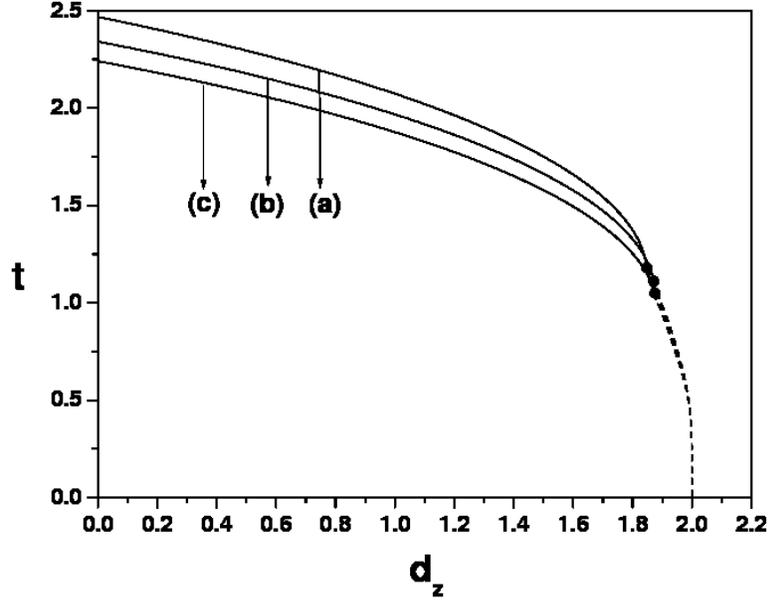}
\caption{Phase Diagrams obtained for  $d_{x}=0$, using clusters with   $N_{c}=1$, 4 and 16 central sites, corresponding to lines
(a), (b) e (c), respectively. The continuous lines represent second-order frontiers, whereas dashed lines are for the first-order ones. The black circles represent tricritical points. } 
\label{figura3}
\end{figure}

\vskip \baselineskip
\vspace{3.0 cm}
\begin{figure}[htbp]
\centering
\includegraphics[height=7.0cm]{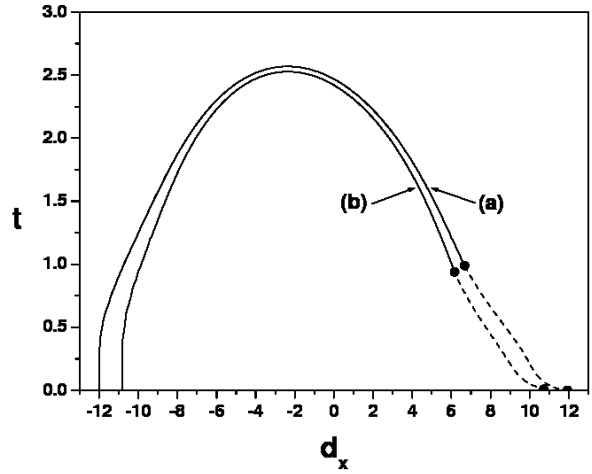}
\caption{  Phase diagrams obtained for  $d_{z}=0$, using clusters of  $N_{c}=1$ and 2 central sites that correspond to lines
(a), (b), respectively. } 
\label{figura4}
\end{figure}
\vskip \baselineskip
\vspace{3.0 cm}
\begin{figure}[htbp]
\centering
\includegraphics[height=8.0cm]{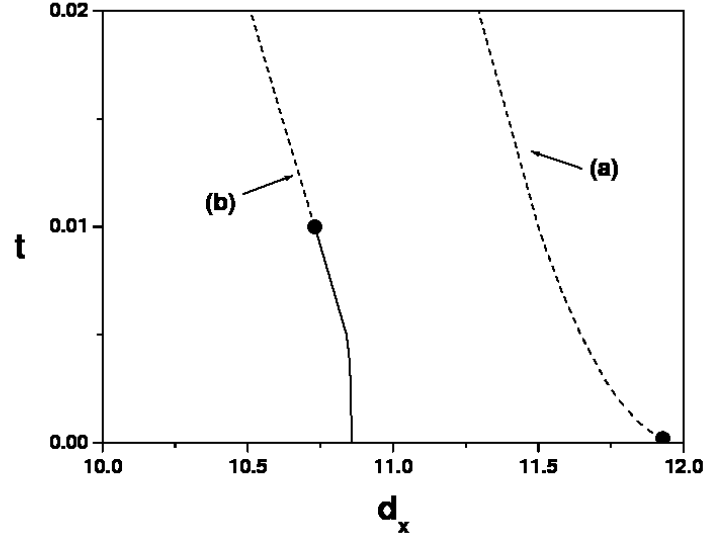}
\caption{ Detailed low temperature region on the right of Fig.4. We remark that lines (a) and (b) correspond to clusters  of  $N_{c}=1$ and 2 central sites, respectively.} 
\label{figura5}
\end{figure}
\vskip \baselineskip
\vspace{3.0 cm}
\begin{figure}[htbp]
\centering
\includegraphics[height=7.0cm]{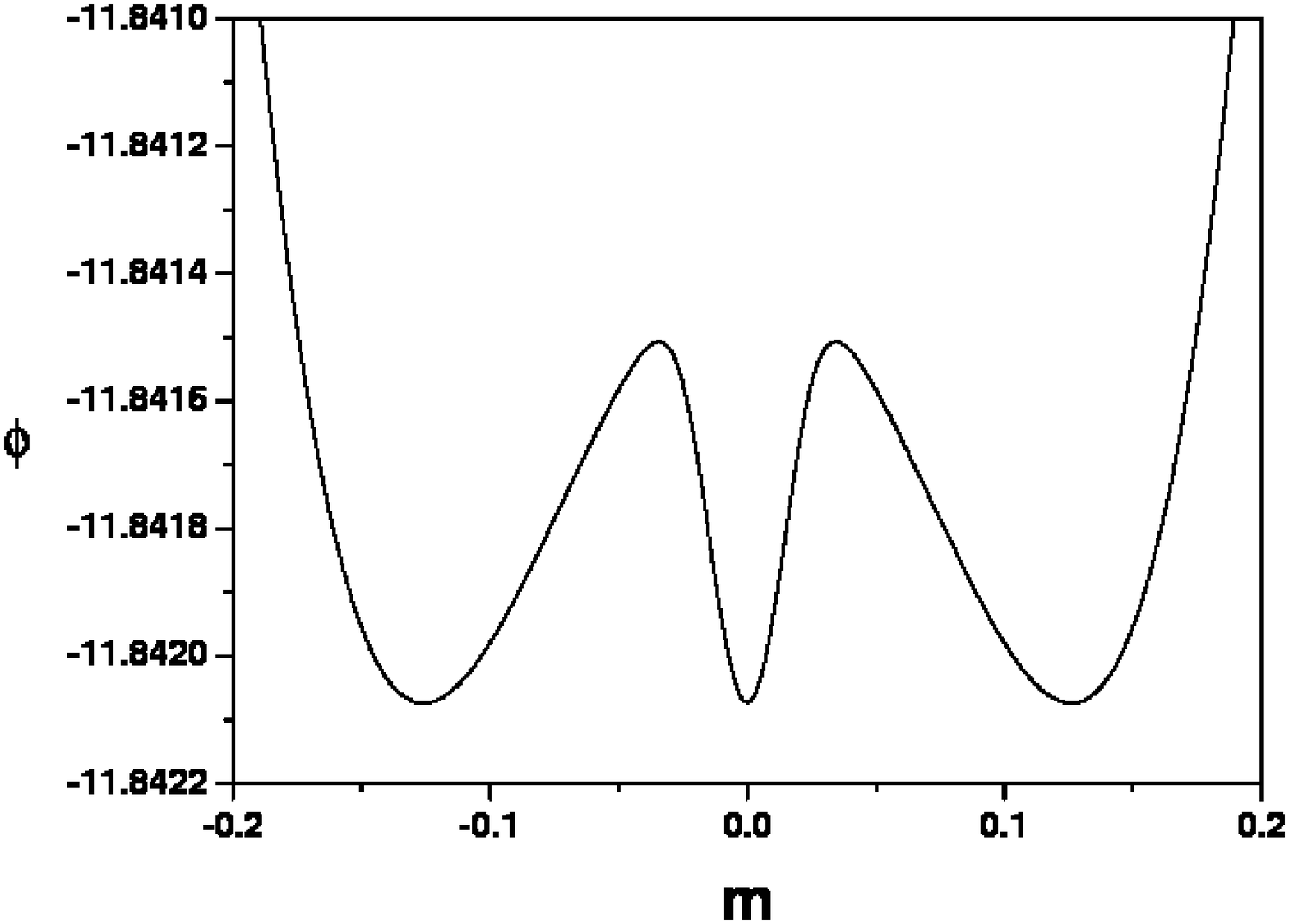}
\caption{Energy minima obtained for the phase transition point $%
P(t=0.0010(1),d_{x}=11.8420(1))$, which corresponds to a first-order point in
Fig.5, for the frontier line corresponding to the size $N_{c}=1$.} 
\label{figura6}
\end{figure}

\vskip \baselineskip
\vspace{3.0 cm}
\begin{figure}[htbp]
\centering
\includegraphics[height=8.0cm]{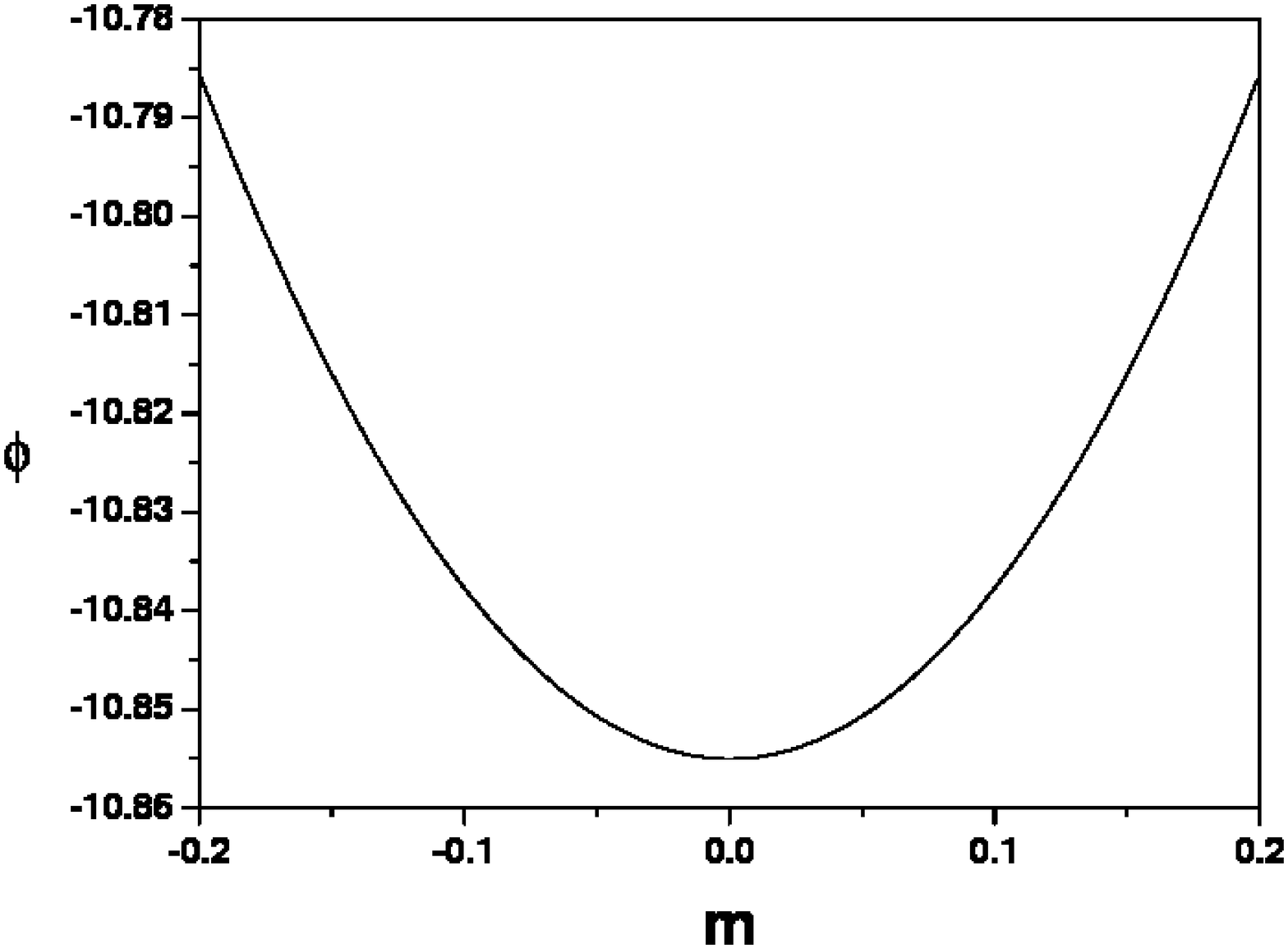}
\caption{ Energy minimum obtained for the phase transition point $P(t=0.0010(1),d_{x}=10.855(1))$, which corresponds to a second-order point in Fig.5,  for the frontier corresponing to the size $N_{c}=2$.} 
\label{figura8}
\end{figure}

\vskip \baselineskip
\vspace{3.0 cm}
\begin{figure}[htbp]
\centering
\includegraphics[height=8.0cm]{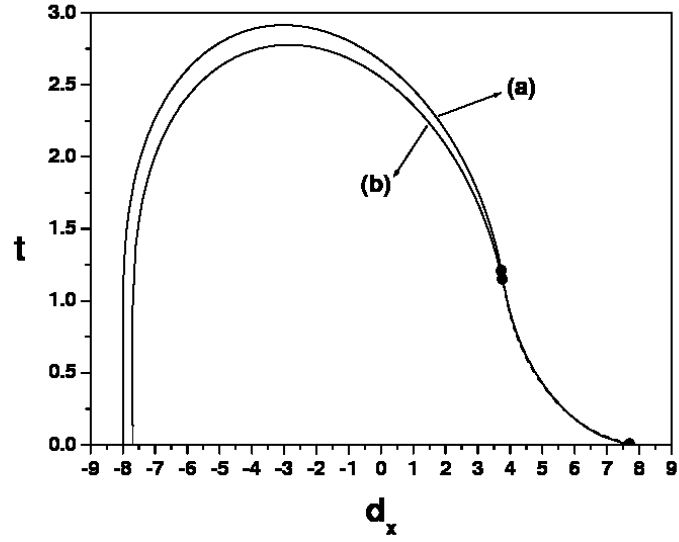}
\caption{Phase diagrams obtained for $d_{z}=0$, using the traditional MFT ($\lambda =1$), where we have in (a) $N_{c}=1$ central sites and in (b) $N_{c}=2$ central sites.} 
\label{figura8}
\end{figure}

\vskip \baselineskip
\vspace{3.0 cm}
\begin{figure}[htbp]
\centering
\includegraphics[height=8.0cm]{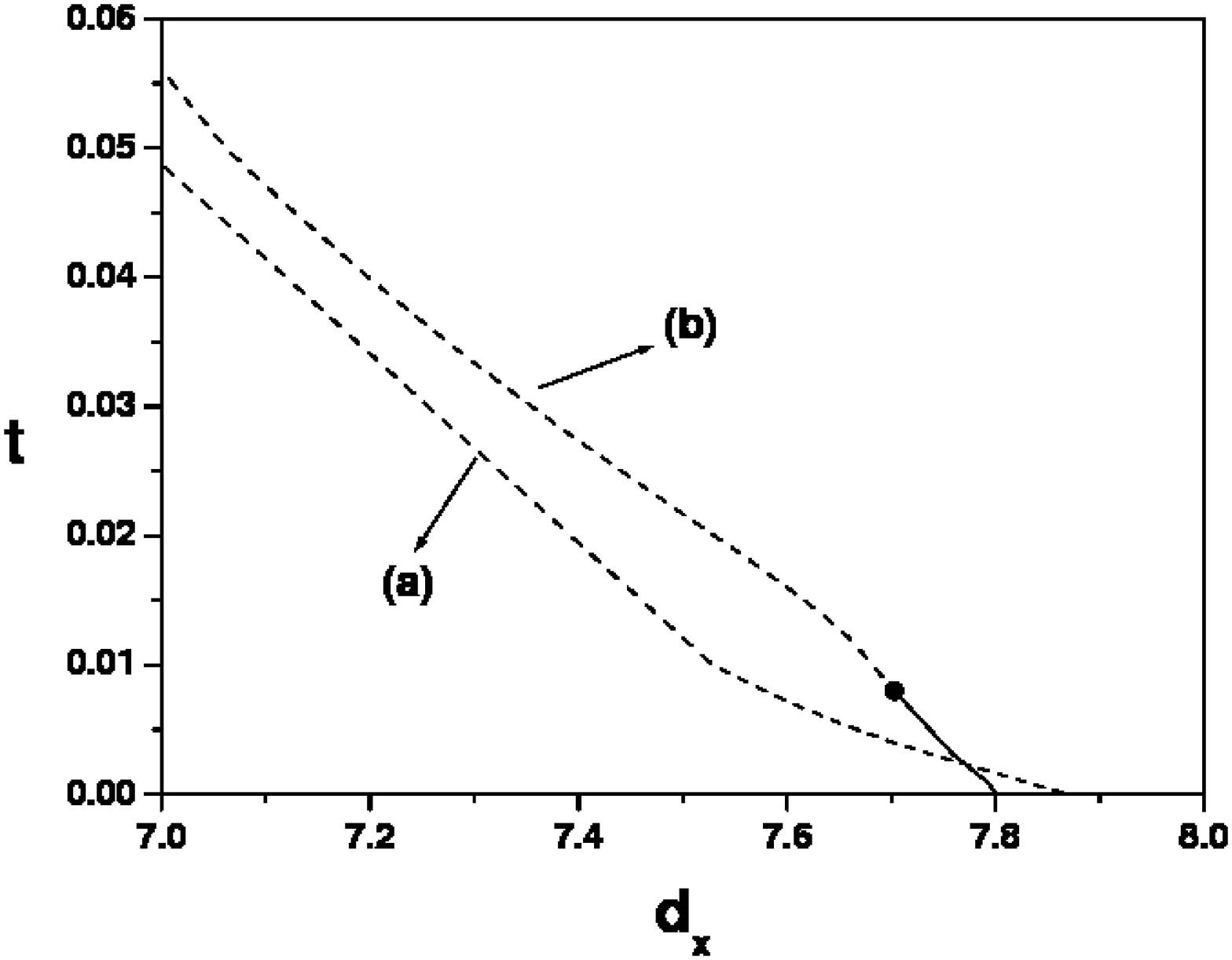}
\caption{Detailed low temperature region on the right of Fig.8. The frontier line (a) ends at zero temperature as a first-order line, whereas the frontier line (b) is of second-order at zero temperature with a tricritical point very close to it. } 
\label{figura8}
\end{figure}

\end{document}